\documentstyle[preprint,aps,epsf]{revtex}
\begin{document}
\begin{titlepage}

\begin{flushright}
   hep-ph/9612356 \\
   MSUHEP--61222  \\
   \date \\
\end{flushright}

\begin{center}
  {\large \bf  Diffractive Photoproduction of $Z^0$ }
  \vskip 0.30in
  {\bf Jon Pumplin }
\vskip 0.1in
Physics and Astronomy Department \\
Michigan State University\\
East Lansing MI 48824, U.S.A.

\end{center}

\vskip 0.5in

\begin{abstract}
The two-gluon exchange model of the pomeron is used to compute
the photoproduction reaction $\gamma p \to Z^0 p$.  The
predicted cross section is too small to be observed at HERA,
but may be detectable at an eventual Next Linear Collider.
\end{abstract}

\end{titlepage}

\section {Introduction}
\label{sec:intro}

The two-gluon exchange model
\cite{lownuss,subtractive,landshoff,cudell,halzen,nikolaev,vmeson}
has for a long time offered a semi-quantitative understanding
of the pomeron in QCD.  The model originated as a view of
elastic scattering.  It has been applied extensively to
vector meson photoproduction, and has recently
been used to predict
jet production in double pomeron interactions \cite{dpe,berera}.
In this paper, we use the model to calculate exclusive $Z^0$
photoproduction at low $Q^2$.

The diagrams for $\gamma p \to Z^0 p$ are shown in Figure~1.
The upper half of each diagram can be calculated in a
straight-forward manner, since only known electroweak and gluonic
couplings appear.  The lower half of each diagram involves the
non-perturbative color structure of the proton, but the required
discontinuity can be obtained by parametrizing it and constraining
the parameters to fit $pp$ elastic scattering.
The calculation of $Z^0$ photoproduction thus introduces no new
parameters.   A similar calculation can be done for
$\gamma p \to \Upsilon p$, where $m_\Upsilon$ provides a large
momentum scale; but that calculation is not as clean because it
brings in the $\Upsilon$ hadronic wave function.

Reggeization of the gluons and interactions between them (e.g., as
given by ladder diagrams) must be responsible for the gradual
energy dependence of pomeron exchange \cite{delduca}.  We will
ignore these effects here.  From a practical standpoint, the energy
dependence of $\gamma p \to Z^0 p$ will be dominated by the minimum
longitudinal momentum transfer that is kinematically required when
the energy is not far above threshold.

The predicted cross section is necessarily small, because
the large mass $m_Z$ sets a short distance scale $\sim 1/m_Z$
for the typical transverse separation between the $q \bar q$ pair
that form the $Z^0$.  This leads to a factor $1/m_Z^2$ in the
final amplitude, which is proportional to the color dipole moment
of the pair.  A couple of enhancement factors $\propto \ln m_Z$
arise in the calculation, as explained below; but the
final predicted cross section turns out to be too small to observe
at the HERA $ep$ collider in its present form.

\section {Calculation of \protect $\gamma p \to Z^0 p$}
\label{sec:photoproduction}
The imaginary part of the amplitude is equal to $1/2i$ times the
discontinuity given by the four diagrams of Figure~1.  The upper
half of diagram (a), for example, corresponds to a factor
\begin{eqnarray}
{\cal D_{\mu \nu}}^{(a)} = \,
\int \frac{d^4 k_1}{(2 \pi)^4} \,
\frac{[2 \pi \delta(k_1^2 - m^2)] \,
[2 \pi \delta(k_2^2 - m^2)]}
{[(p_1 - k_1)^2 - m^2] \,
 [(p_3 - k_2)^2 - m^2]} \; {\rm tr}_a
\label{eq:eq1}
\end{eqnarray}
in the discontinuity, where
\begin{eqnarray}
{\rm tr}_a &=& \,
Tr\{\gamma \! \cdot \! \epsilon(p_1) \,
[m - i \gamma \! \cdot \!(k_1 - p_1)] \,
\gamma_\mu \,
(m + i \gamma \! \cdot \! k_2) \,
(A + B \gamma_5) \,  \nonumber \\
& & \; \; \quad \gamma \! \cdot \! \epsilon^*(p_3) \,
[m - i \gamma \! \cdot \!(k_1 - q_2)] \,
\gamma_\nu \,
(m - i \gamma \! \cdot \!k_1) \} \,
 \; .
\label{eq:eq2}
\end{eqnarray}
To obtain the high energy limit of the amplitude, we use
light-cone
coordinates $p_\pm = (p_0 \pm p_z)/\sqrt{2}$ where $p_{1+}$ and
$p_{2-}$ are large and $s \cong p_{1+} \,p_{2-} \,$.
One of the two delta functions in Eq.~(\ref{eq:eq1}) is saved
for the eventual integration over gluon momentum $q_1$, which
carries no large $+$ or $-$ component as a result of this and a
similar delta function from the lower half of the diagram.
The other delta function reduces Eq.~(\ref{eq:eq1}) to a
three-dimensional integral over transverse momentum and
light-cone momentum fraction
$x = k_{1+}/p_{1+}$:
\begin{eqnarray}
{\cal D_{\alpha \beta}}^{(a)} &=& \,
\delta(p_1 \! \cdot \! q_1 + {\cal O}(1))
\, p_{1 \alpha} \, p_{1 \beta} \, T_a \,  \nonumber \\
\, \null \nonumber \\  
T_a &=&
\frac{1}{4 \pi^2 s^2} \,
\int_0^1 \frac{dx}{x \, (1-x)} \,
\int d^2 k_{1\perp} \, \frac{p_{2 \mu} \, p_{2 \nu} \,
{\rm tr}_a }{d_1 \, d_2}
\; ,
\label{eq:eq3}
\end{eqnarray}
where
\begin{eqnarray}
d_1 &=& (p_1 - k_1)^2 - m^2 \cong
[(k_1 - x p_1)_\perp^2 + m^2 - x(1-x)p_1^2] / x  \,  \nonumber \\
d_2 &=& (p_3 - k_2)^2 - m^2 \cong
[(k_1 - q_2 - x p_3)_\perp^2 + m^2 - x(1-x)p_3^2] / (1 - x)
\; .
\label{eq:eq4}
\end{eqnarray}
For $Q^2 = 0$ we can choose a gauge in which $\epsilon(p_1)$ has
only transverse
components.  Then $\epsilon(p_3)$ can be assumed to have only
transverse components as well, because the production of
longitudinal $Z^0$ can be shown to be suppressed by a factor
$\sqrt{-t}/m_Z$ in the amplitude.

Upon summing over all four diagrams, the contribution from
$\gamma_5$ terms is found to vanish in the large $s$ limit.
Combining the denominators in Eq.~(\ref{eq:eq3}) using the
Feynman parameter trick
$\frac{1}{a\,b} = \int_0^1 dy \, [ay + b(1-y)]^{-2}$ allows
the $k_{1\perp}$ integral to be performed, and leads to
\begin{eqnarray}
{\cal D_{\alpha \beta}}^{(1)} &=& \,
\delta(p_1 \! \cdot \! q_1 + {\cal O}(1))
\, p_{1 \alpha} \, p_{1 \beta} \, T^{(1)}  \label{eq:eq5} \\
\, \null \nonumber \\  
T^{(1)} &=&
F^{(1)}(q_{2 \perp}, \Delta_\perp) - F^{(1)}(0, \Delta_\perp)
\label{eq:eq6}
\end{eqnarray}
where
\begin{eqnarray}
F^{(1)}(q_\perp, \Delta_\perp) &=&
\frac{A}{\pi} \, \int_0^1 dx \int_0^1 dy \,
\{
\epsilon_\perp(p_1) \! \cdot \! \epsilon^*_\perp(p_3) \,
( [m^2 - y(1-y)v^2] / c \nonumber \\
 &-&  [1 - 2x(1-x)] \ln c )
+ 4 \, v \! \cdot \! \epsilon_\perp(p_1)  \,
       v \! \cdot \! \epsilon_\perp^*(p_3) \,
x \, (1-x)\, y\, (1-y)/ c \}  \nonumber \\
\, \null \nonumber \\  
c &=& m^2 + y \, (1-y) \, v^2 - x \, (1-x) \,
[y \, p_3^2 + (1-y) \, p_1^2] \nonumber \\
\, \null \nonumber \\  
v &=& q_\perp + x \Delta_\perp \; .
\label{eq:eq7}
\end{eqnarray}
Here $p_1^2 = -Q^2$ is approximately zero,
$p_3^2 = m_Z^2$, and
$\Delta_\perp = (p_3 - p_1)_\perp  = (q_1 - q_2)_\perp$
is the transverse momentum transfer.

The $F(q_{2 \perp}, \Delta_\perp)$ term in Eq.~(\ref{eq:eq6})
comes from the two ``off-diagonal'' diagrams in which one gluon
is attached to the quark line and the other to the anti-quark.
The $F(0, \Delta_\perp)$ term comes from the diagonal diagrams
in which both gluons are attached to the same line.
There is a strong
cancellation between these two contributions when
$q_{1 \perp}$ or $q_{2 \perp}$ is small, because the
two-gluon system is color neutral and the $q \bar q$ system is
spatially compact due to the large $Z^0$ mass.  However,
the diagonal term will dominate in the full amplitude because
the range of eventual integration over $q_{\perp}$ is controlled
by $m_Z$.

We can estimate the integral in Eq.~(\ref{eq:eq7}) using the
approximation
$m_Z^2 \gg q_{1 \perp}^2$, $q_{2 \perp}^2$, $\Delta_\perp^2$.
This limit is obtained by splitting the integration into
six separate regions according to
$0 < y < y_0$ or $y_0 < y < 1$ and
$0 < x < x_0$, $x_0 < x < 1 - x_0$, or $1 - x_0 < x < 1$.
The result is independent
of $x_0$ and $y_0$ in the limit $x_0 \ll 1$, $y_0 \ll 1$
and is given by
\begin{eqnarray}
T^{(1)} &=&
\frac{-A m^2}{\pi m_Z^2} \,
\{\epsilon_\perp(p_1) \! \cdot \! \epsilon^*_\perp(p_3) \,
[f(q_{1\perp}^2 / m^2)
 + f(q_{1\perp}^2 / m^2)
 - f(\Delta_\perp^2 / m^2)] \nonumber \\
& &
\, + \,
[q_{1\perp} \! \cdot \! \epsilon_\perp(p_1) \,
 q_{2\perp} \! \cdot \! \epsilon^*_\perp(p_3)
\, + \,
 q_{2\perp} \! \cdot \! \epsilon_\perp(p_1) \,
 q_{1\perp} \! \cdot \! \epsilon^*_\perp(p_3)] / m^2 \}
\nonumber \\
\, \null \nonumber \\  
f(a) &=& a \, [ \ln (m^2/m_Z^2) - i\pi - \frac{1}{2} +
b \ln \frac{b+1}{b-1} \, ] \nonumber \\
\, \null \nonumber \\  
b &=& \sqrt{1 + 4/a}
\label{eq:eq8}
\end{eqnarray}

We want the limit of high energy at small momentum
transfer, where the helicity non-flip amplitude dominates.
We can therefore take
$\epsilon_\perp(p_3) \cong \epsilon_\perp(p_1) =
\frac{1}{\sqrt{2}}(\mp 1, -i, 0, 0)$ to obtain
\begin{eqnarray}
& &\epsilon_\perp(p_1) \! \cdot \! \epsilon^*_\perp(p_3)
\cong 1 \, , \nonumber \\
\, \null \nonumber \\  
& &q_{1\perp} \! \cdot \! \epsilon_\perp(p_1) \,
q_{2\perp} \! \cdot \! \epsilon^*_\perp(p_3)
\, + \,
q_{2\perp} \! \cdot \! \epsilon_\perp(p_1) \,
q_{1\perp} \! \cdot \! \epsilon^*_\perp(p_3)
\cong
q_{1\perp} \! \cdot \! q_{2\perp} \; .
\label{eq:eq9}
\end{eqnarray}
It is a good approximation to neglect the quark mass $m$, since
Eq.~(\ref{eq:eq8}) is not singular in the limit $m \! \to \! 0$.
This approximation is reasonably good numerically
even for quark masses up to the charm quark mass, in the
important regions of $q_{1\perp}^2$, $q_{2\perp}^2$ and
$\Delta^2$, because the
corrections to it do not contain the large
factor $\ln m_Z^2$.  In this way, we find
\begin{eqnarray}
T^{(1)} &\cong&
\frac{-A}{\pi \, m_Z^2} \,
[g(q_{1\perp}^2) + g(q_{2\perp}^2) - g(\Delta_\perp^2)]
\nonumber \\
g(q^2) &=& q^2 \, [\ln q^2 / m_Z^2 - i\pi] \; .
\label{eq:eq10}
\end{eqnarray}
Eq.~(\ref{eq:eq10}) was derived for the region where
$q_{1 \perp}^2$, $q_{2 \perp}^2$, and
$\Delta_\perp^2 = (q_1 - q_2)_\perp^2$ are small compared to
$m_Z^2$.   We can provide it with approximately correct
behavior when these quantities become comparable to $m_Z^2$ by
inserting an additional factor
$[1 + (q_{1\perp}^2 + q_{2\perp}^2)/m_Z^2]^{-1}$ into
Eq.~(\ref{eq:eq10}).  The adequacy of the resulting approximation
to $T^{(1)}$, for the purpose of calculating the full amplitude,
has been checked by numerical integration.

The coefficient $A$ in Eq.~(\ref{eq:eq10}) is given by
\begin{eqnarray}
A =
\frac
{4 \, \pi^2 \, \alpha \, \alpha_s}
{\sin \theta_w \cos \theta_w} \,
(1 - \frac{20}{9} \, \sin^2 \theta_w) \; .
\label{eq:eq11}
\end{eqnarray}
This includes a factor $2$ from the sum over quark flavors
$u + d$ and $s + c$, and a factor $1/2$ from the sum over quark
colors.  I take $\alpha_s \cong 0.25$ for the strong coupling
at the low average momentum transfer scale that occurs here.
This is slightly optimistic, since the amplitude receives
contributions from a spectrum of gluon transverse momenta
extending all the way up to $\sim m_Z$.  It leads to $A = 0.083 \,$.

A reasonable parametrization for the discontinuity of the gluon-proton
amplitude represented by the bottom half of Figure~1 has been
given in  Ref. \cite{dpe}:
\begin{eqnarray}
{\cal D_{\mu \nu}} &=& \,
\delta(p_2 \! \cdot \! q_1 + {\cal O}(1))
\; p_{2 \mu} \, p_{2 \nu} \, T^{(2)}
\label{eq:eq12}
\end{eqnarray}
where
\begin{eqnarray}
T^{(2)} &=& F^{(2)}(q_{2 \perp}, \Delta_\perp) -
F^{(2)}(0, \Delta_\perp) \nonumber \\
\, \null \nonumber \\  
F^{(2)}(q_{\perp}, \Delta_\perp) &=&
N \, \sqrt{\frac{2 \pi a b}{\beta}} \;
(a+b)^{-2} \;
 e^{-(\beta/2) \, [\, (a+b)^2 - \Delta_\perp^2 \, ]} \; ,
\label{eq:eq13}
\end{eqnarray}
with
$a = \sqrt{q_{\perp}^{\, 2} + 4m_q^2}$ and
$b = \sqrt{(\Delta - q)_\perp^{\, 2} + 4m_q^2} \,$.
A selection of reasonable parameter choices that have been tuned
to fit $pp$ elastic scattering is given in Table~I.  Dependence of
the results on the choice of parameter set provides an estimate
of systematic errors.
\begin{table}
\begin{center}Table I \\
Parameters and predicted cross section
$\sigma$ and slope $B$ for $\gamma p \to Z^0 p$.
\end{center}
\begin{center}
\begin{tabular}{||c|c|c|c||c|c||}
\hline
\multicolumn{1}{||c|}{$m_g$} &
\multicolumn{1}{c|}{$m_q$} &
\multicolumn{1}{c|}{$\beta$} &
\multicolumn{1}{c||}{$N$} &
\multicolumn{1}{c|}{$\sigma [{\rm pb}]$} &
\multicolumn{1}{c||}{$B [{\rm GeV}^{-2}]$} \\
\hline
\hline
$0.14$ & $0.5$ &
$5.797$ & $2.65 \times 10^7$ &
$0.014 $ & 7.0 \\

$0.14$ & $0.3$ &
$4.944$ & $4.89 \times 10^3$ &
$0.017 $ & 6.4 \\
\hline
$0.30$ & $0.5$ &
$6.659$ & $2.13 \times 10^8$ &
$0.018 $ & 7.5 \\

$0.30$ & $0.3$ &
$5.901$ & $1.40 \times 10^4$ &
$0.020 $ & 6.9 \\
\hline

$1.0$ & $0.5$ &
$7.638$ & $4.06 \times 10^9$ &
$0.045 $ & 7.9 \\

$1.0$ & $0.3$ &
$7.098$ & $9.04 \times 10^4$ &
$0.047 $ & 7.5 \\
\hline
\hline
\end{tabular}
\label{table1}
\end{center}
\end{table}

The full amplitude for $\gamma p \to Z^0 p$ in the two gluon
exchange picture of Figure~1 (including a factor $8$ from the sum
over gluon colors) is given by
\begin{eqnarray}
{\cal M} = \, \frac{i \, s}{8 \, \pi^4} \,
\int  \,
\frac{d^2 q_\perp \;
T^{(1)}(q_\perp, \Delta_\perp) \;
T^{(2)}(q_\perp, \Delta_\perp) }
{(q_\perp^{\, 2} + m_g^{\, 2}) \,
[(\Delta_\perp - q_\perp)^{\, 2} + m_g^{\, 2}]}
\label{eq:eq14}
\end{eqnarray}
with $\Delta_\perp^{\, 2} = -(p_1 - p_3)^2 = -t$.
A finite gluon mass $m_g$ is included in the propagators
to suppress contributions from long distance, as an
approximation to color confinement \cite{lownuss,subtractive}.
In elastic scattering, this is necessary to avoid unphysical
behavior of the elastic slope in the limit $t \to 0$.
The $q_\perp$ integral
can be done numerically at each momentum transfer
$\Delta_\perp$.  The result is enhanced by two powers of
$\ln m_Z^2$:  one coming directly from $T^{(1)}$
(see Eq.~(\ref{eq:eq10}))
and the other coming from the integration over $q_\perp$,
since the integrand
behaves as $d q_\perp^2 / q_\perp^2$ for large $q_\perp^2$ up
to $\sim m_Z^2)$.

The differential cross section is given by
\begin{eqnarray}
\frac{d\sigma}{dt} = \frac{1}{16 \, \pi \, s^2} |{\cal M}|^2
\label{eq:eq15}
\end{eqnarray}
in the high energy limit.  It is independent of $s$ in that
limit.  The predicted integrated cross section
$\sigma = \int \frac{d\sigma}{dt} \, dt$ is listed in
Table~I for each choice of parameters.  The result is
approximately $0.025 \, {\rm pb}$, with an uncertainty of about
a factor of 2 coming from the range of parameters that provide a
plausible description of elastic scattering.  (It is interesting
to note that this calculated result is not far from
the crude estimate of $\approx 0.08 \, {\rm pb}$ that can be
made by scaling the observed cross section for
$\gamma \, p \to J/\psi \, p$ at HERA \cite{psi} by the dipole
size factor $(m_{J/\psi}/m_Z)^4$.)

The dependence of $\frac{d\sigma}{dt}$ on momentum transfer
$t$ is roughly exponential, so we can characterize it by a
slope parameter $B$ such that
$\frac{d\sigma}{dt} \propto e^{B \, t}$.  The value of $B$,
defined by reproducing $\sigma$ and $\frac{d\sigma}{dt}$
at $t=0$, is also listed in Table~I.  It is approximately
equal to the one half the slope of elastic scattering,
since it arises essentially from the wave function effect
associated with the proton remaining intact, while in $pp$
elastic scattering, {\it both} protons must remain intact.
Equivalently, one may say that the extent of the interaction
in impact parameter space is set by that of the proton.

\section {Prediction for \protect $e p \to e Z^0 p$}
\label{sec:hera}

One can hope to look for $Z^0$ photoproduction in
electroproduction $e p \to e Z^0 p$, which will
be dominated by low $Q^2$ transverse photons producing $Z^0$'s
with the same helicity as the photon.  The cross section
is given by
\begin{eqnarray}
\frac{d\sigma(e \, p \to e \, Z^0 \, p)}{dy \, dQ^2} =
f_{\gamma/e}(y,Q^2) \; \sigma_{\gamma^* \, p}(W) \; ,
\label{eq:eq16}
\end{eqnarray}
where
\begin{eqnarray}
f_{\gamma/e}(y,Q^2) =
\frac{\alpha}{2 \, \pi \, Q^2} \left[
\frac{1 + (1 -y)^2}{y} - \frac{2 \, m_e^2 \, y}{Q^2} \right]
\label{eq:eq17}
\end{eqnarray}
is the flux of transverse photons \cite{psi}.
Defining the four-momenta in $e \, p \to e \, Z^0 \, p$ as
$k$, $p$, $k^\prime$, $q^\prime$, $p^\prime$ respectively,
the usual kinematic variables are
$s = (k + p)^2$,
$Q^2 = -(k^\prime - k)^2$, and
$y = (k - k^\prime)  \cdot  p / k \cdot p \,$.
The role of $s$ in Sect.\ \ref{sec:photoproduction} is now played by
$W^2 = (k - k^\prime + p)^2 \cong y s$.
The minimum value of $Q^2$ is given by
$Q_{\rm min}^2 = m_e^2 \, y^2 / (1-y)$.  The maximum $Q^2$ occurs
when the electron is scattered through $180^\circ$, but most of the
cross section comes from the region of very small $Q^2$.
(We have assumed the limit $Q^2 \to 0$ in calculating the cross
section for $\gamma p \to Z^0 p$ in Sect.\ \ref{sec:photoproduction},
but if that condition is relaxed, the model shows no dramatic
dependence on $Q^2$.)

We assume that the
dependence on $W^2$ is dominated by the effect of the minimum
momentum transfer $|t_{\rm min}|$.  Specifically, we assume
the approximate dependence $\sim e^{B \, t}$ with
$B \approx 7 \, {\rm GeV^{-2}}$, which was found in
the large $W$ limit, to be approximately valid at finite $W$
where the minimum value of $t = (p - p^\prime)^2$ is
different
from $0$ because of the longitudinal momentum transfer
necessary to produce the high-mass state.
This assumption might appear optimistic because producing
the on-shell intermediate states in Figure~1 involves an
increase in mass of the states at both ends of each gluon line,
which requires a noticeably larger minimum longitudinal momentum
transfer.  However, the scale of transverse momentum transfer
for each gluon exchange extends all the way to ${\cal O}(m_Z)$,
so the longitudinal momentum transfer associated with the
individual gluon exchanges need not be dominant.

Integrating Eq.~(\ref{eq:eq16}) at the HERA energy
$\sqrt{s} = 300 \, {\rm GeV}$ leads to
$\sigma_{e p \to e Z^0 p} / \sigma_{\gamma p \to Z^0 p} = 0.032$.
Combining this with the results listed in Table~I yields a
predicted cross section for diffractive $Z^0$ production at HERA
of approximately 1 femtobarn.   A realistic experimental search
would have to rely on one of the clean decay modes
$Z^0 \to e^+e^-$ or
$Z^0 \to \mu^+ \mu^-$,
which would suppress the rate by a branching fraction $0.067$.
The resulting predicted rate is too small by 2 -- 3 orders of
magnitude to be observed at the HERA $ep$ collider in its present
form.

\section{ Conclusion}
\label{sec:conclusion}

The quasi-elastic photoproduction process $\gamma p \to Z^0 p$ has
been calculated using the two-gluon exchange model of the pomeron.
The calculation is rather clean because the electroweak part of
it is completely well defined and the soft proton part can
be normalized to elastic scattering.

The process would lead to $Z^0$ production in $ep$ collisions,
which would have a clean and dramatic signature: one
$e^+ e^-$ or $\mu^+ \mu^-$ pair with invariant mass equal to
the $Z^0$ mass, and hence with very large individual transverse
momenta; and nothing else in the detector, since the
scattered electron and proton would generally disappear down
the beam pipes.  Unfortunately, the predicted rate is too small
to be observed with the luminosity currently available at HERA.
Possible upgrades to HERA that would increase its luminosity by
2 --3 orders of magnitude could make the process observable.

Because of the higher available energy, the related diffractive
process $\gamma \gamma \to Z^0 \rho^0$ should be observable
with a $\gamma \gamma$ option at a ${\cal O}(1 \, {\rm TeV})$
Next Linear Collider \cite{peskin}.  The cross section can be
estimated from Sect.\ \ref{sec:photoproduction} and Regge
factorization as $\sim \! 0.3 \, {\rm femtobarns}$ for
$\gamma \gamma$ energies far above threshold.

\section*{Acknowledgments}
This work was
supported in part by U.S. National Science Foundation grant
number PHY-9507683.


\begin{figure}
    \begin{center}
       \leavevmode
       \epsfxsize=0.8\hsize
      \epsfbox{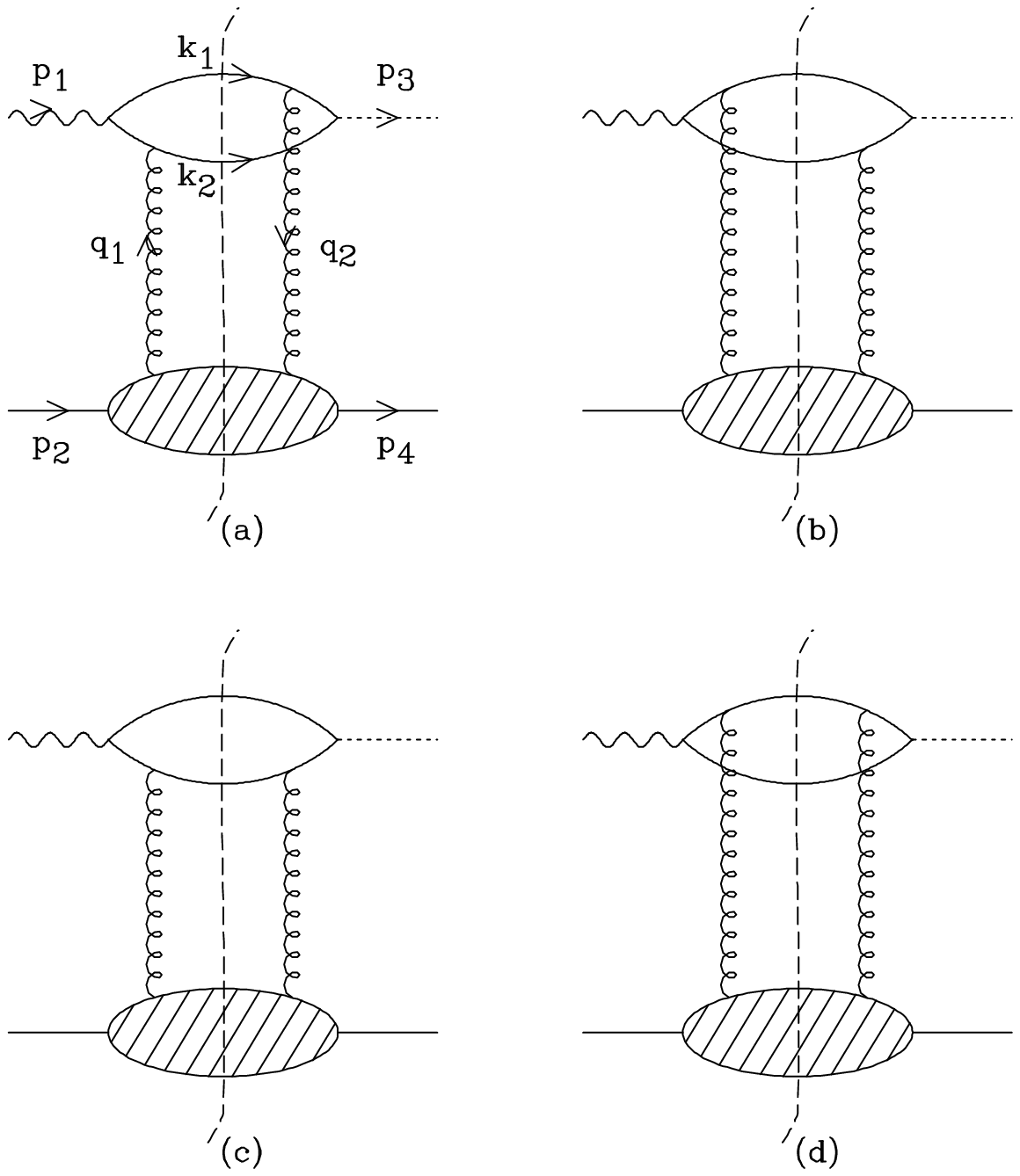}
    \end{center}
\caption{\protect
Two gluon exchange model for $\gamma p \to Z^0 p$.
The dashed line denotes an $s$-channel discontinuity.
}
\label{figure1}
\end{figure}
\end{document}